\documentclass{aa}
\usepackage{graphicx}

\usepackage{natbib}

\bibpunct{(}{)}{;}{a}{}{,} 

\newcommand{\RM}{\mathrm{RM}}
\newcommand{\DM}{\mathrm{DM}}
\newcommand{\HH}{B}
\newcommand{\nel}{n_\mathrm{e}}

\newcommand{\half}{{\textstyle{1\over2}}}

\begin{document}
   \title{Wavelet tomography of the Galactic magnetic field}

   \subtitle{I. The method}

\author{R.~Stepanov,$^1$ P.~Frick,$^1$  A.~Shukurov$^{2}$ and
D.~Sokoloff$\,^{3}$}

   \offprints{R.~Stepanov, \email{rodion@icmm.ru}}

   \institute{
   $^1$Institute of Continuous Media Mechanics,
Korolyov str.~1, 614061 Perm, Russia \\
$^2$Department of Mathematics, University of Newcastle, Newcastle
upon Tyne NE1~7RU, U.K.\\
$^3$Department of Physics, Moscow University, 119899, Moscow,
Russia}

\titlerunning{Wavelet tomography}

\authorrunning{R.~Stepanov et al.}

   \date{Received ... / Accepted ...}

   \abstract{We suggest a two-dimensional wavelet devised to
deduce the large-scale structure of a physical field (e.g., the
Galactic magnetic field) from its integrals along straight paths
from irregularly spaced data points to a fixed interior point (the
observer). The method can be applied to the analysis of pulsar
rotation and dispersion measures in terms of the large-scale
Galactic magnetic field and electron density. The method does not
use any \emph{\'a priori\/} assumptions about the physical field
and can be considered as an algorithm of wavelet differentiation.
We argue that a certain combination of the wavelet transformation
with model fitting would be most efficient in the interpretation
of the available pulsar $\RM$ data.
   }
\maketitle

\section{Introduction}

Tomography, understood broadly, is a reconstruction of a
multidimensional physical field from its integral projections
obtained by exposing it in different aspects. Tomography problems
often occur in astronomy, especially radio astronomy, where the
intervening medium is transparent and the observable quantities
represent certain integrals along the line of sight.  An important
example is the studies of the magneto-ionic medium in the Milky
Way using the Faraday rotation measure $\RM$ and dispersion
measure $\DM$ of pulsars -- both are integrals along the line of
sight involving interstellar magnetic field and thermal electron
density. A peculiar feature of the astronomical tomography is that
just one vantage point is available, as the observer is confined
to the close vicinity of the Sun. Astronomical tomography then
focuses on extended objects which can be meaningfully analyzed
using the integral projections obtained in a variety of
directions.

In this paper we suggest a method to assess the spatial structure
of the global Galactic magnetic field using pulsar $\RM$ (and also
$\DM$ as the Faraday rotation measure depends on the thermal
electron density), one of the most informative tracers of the
large-scale component of the magnetic field \citep[see
e.g.,][]{Beck2000}. A fundamental problem here is that the
derivation of magnetic field $\vec{\HH}$ from the observable
integral
\begin{equation}
\RM \equiv K\int_0^r \nel {\vec{\HH}}\cdot d{\vec{s}}\;,
\label{RM}
\end{equation}
in fact involves differentiation of the observed quantity with
respect to $r$ resulting in a catastrophic amplification of noise.
Moreover, the distances to pulsars, $r$, are contaminates by large
errors, both random and systematic. The problem is further
aggravated by the fact that the pulsars (and extragalactic radio
sources), that provide the $\RM$ data, are irregularly spaced and
cover the sky very inhomogeneously. Here $\nel$ is the number
density of thermal electrons, $r$ is the distance to the radio
source and $K=0.81\,{\rm rad\,m^{-2}\,cm^3\,\mu G^{-1}\,pc^{-1}}$;
the positive direction of $\vec{s}$ is that towards the observer,
so that a field directed towards the observer produces positive
$\RM$. The thermal electron density can be obtained from the
dispersion measure of pulsars \citep{manch72,manch74}:
\begin{equation}
\DM\equiv\int_0^r \nel \,d s\,, \label{DM}
\end{equation}
and this involves similar complications.

The idea of this paper is twofold. Firstly, we use wavelets to
filter out the noise in the $\RM$ data resulting from small-scale
fluctuations in the magneto-ionic medium and from the irregular
spacing of the data points. Secondly, we introduce a new,
specialized family of wavelets devised to avoid (or at least to
minimize) noise amplification in dealing with an observable
represented by an integral along the line of sight, given that the
available data probe a range of distances into the field
localization region. The Faraday rotation and dispersion measures
of pulsars are suitable observables for such an analysis.

The plan of the paper is as follows. In Sect.~\ref{data_s} we
discuss the available samples of pulsar $\RM$ and $\DM$ and
methods used to obtain $\vec{B}$ and $\nel$ from them.
Section~\ref{wl_s} introduces the basic ideas of our method, which
can be applied in various contexts. The algorithm is described in
Sect.~\ref{method_s}, and the advantages and limitations of the
method are discussed in Sect.~\ref{demo_s}. The efficiency of the
method when applied to the pulsar $\RM$ is discussed in
Sect.~\ref{diss_s}.

\section{Obtaining magnetic field from RM and electron density
from DM} \label{data_s}

Faraday rotation measures of pulsars probe the Galactic magnetic
field in a variety of directions. A pulsar's own contribution to
the observed $\RM$ is minor because the pulsar magnetosphere is
populated by electron-positron pairs resulting in zero net Faraday
rotation. This makes pulsars a major source of information about
the large-scale magnetic field of the Milky Way
\citep[e.g.,][]{rk89,rl94}, especially in conjunction with their
dispersion measures.  However, there are several complications in
the derivation of magnetic field and electron density from pulsar
$\RM$ and DM.

   \begin{figure}
   \centering
   \includegraphics{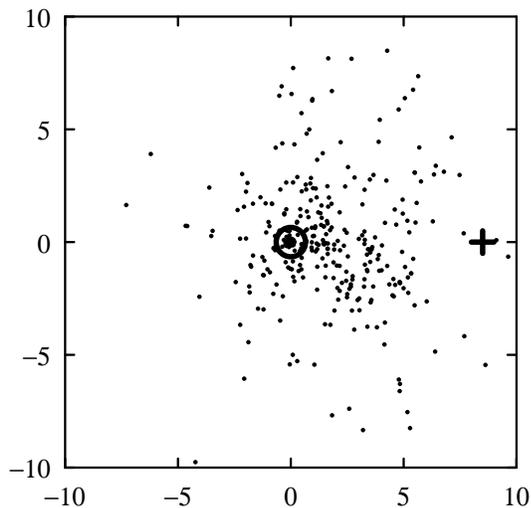}
   \caption{The 323 pulsars with known $\RM$ from the catalogue of \citet{tay95}
are shown in the galactic plane. Here and below, scale is given in
kiloparsecs and the Galactic centre is shown with cross; the Sun
is at the centre of the frame. }
         \label{sampl}
   \end{figure}

Firstly, pulsars are very non-uniformly distributed in space. They
concentrate in spiral arms and the number density of known pulsars
decreases rapidly with distance from the Sun as shown in
Fig.~\ref{sampl}. Even though the number of known pulsars rapidly
grows with time, the non-uniform nature of their spatial
distribution remains. Most pulsars are found near the Galactic
midplane. Yet, some of them are located far away from the
midplane, and thereby allow a study of the vertical distribution
of the magneto-ionic medium. Nevertheless, in this paper we
restrict ourselves to a two-dimensional analysis by projecting the
pulsar data onto the galactic plane; our method can be generalized
to three dimensions, but better data would be needed to justify
this.

   \begin{figure}
   \centering
   \includegraphics{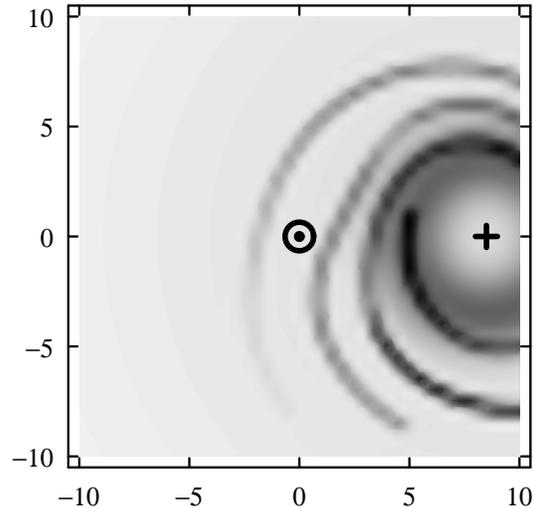}
      \caption{The thermal electron density $\nel$ distribution in the Galactic plane,
      following the model of \protect\citet{1993ApJ...411..674T},
with the Galactic centre marked with cross and the Sun at the
centre of the frame. Darker shades of grey correspond to larger
densities.}
         \label{ne}
   \end{figure}

Secondly, the electron density can be obtained from dispersion
measures only if the distance to the pulsars is known. Distance
estimates now exist for a few hundreds of pulsars, resulting from
three basic techniques: neutral hydrogen absorption (in
combination with the Galactic rotation curve), trigonometric
parallax and from associations with objects of known distance
\citep{pulsreview}. The distance of a pulsar can be obtained from
$\DM$ if the distribution of  $\nel$ is known.
\cite{1993ApJ...411..674T}, based mainly on the scintillation and
dispersion measure data, have suggested an electron density model
expected to provide distance estimates with an average uncertainty
of about $30\%$. However, distances to individual pulsars may be
as wrong as by as a factor of two \citep[e.g.,][]{jkww2001}.

The method suggested here can be applied to pulsar dispersion
measures to obtain thermal electron density using pulsars with
known distances. Then magnetic field can be obtained from $\RM$
using the same method. However, dispersion measures alone are not
sufficient to obtain a reliable distribution of $\nel$, and a
model for the electron density should be used in order to derive
magnetic field distribution from Faraday rotation measures.

The largest catalogue of pulsars contains 706 objects
\citep{tay95}. Faraday rotation and dispersion measures, together
with the distances are known for only 323 of them.

The simplest estimate of the line of sight component of the
large-scale magnetic field $\vec{\HH}$ can be obtained by dividing
$\RM$ by $\DM$. This would be justified if both $\vec{\HH}$ and
$\nel$ were uniform along the line of sight, which is definitely
not the case. A more elaborate analysis involves a pair of pulsars
located close to each other in the sky, so $\nel$ between the two
pulsars can be obtained from the increments in $\DM$ and $r$ as
$\nel\simeq\Delta\DM/\Delta r$. Then the longitudinal magnetic
field follows as $\HH_{||}\simeq\Delta\RM/(K\Delta\nel)$. However,
$\Delta r$ is most often too large to make the resulting estimate
reliable. The nature of the problem is intrinsically related to
that arising in differentiating unevenly sampled data \citep[cf.][
p.\ 34]{rss}. A possible resolution is to use multivariate
statistical analysis where a model for magnetic field is adopted
and its parameters are obtained by fitting \citep{rss,rk89,rl94}.
A disadvantage of this (and any other) analysis of this kind is
that one has to impose \emph{\'a priori\/} model restrictions on
the structure of the magnetic field. Furthermore, it is often
difficult, if not impossible, to get convincing evidence that the
model adopted is compatible with observations because errors in
the observable quantities are rarely known confidently and this
hampers the application of statistical tests for the quality of
the fit. Here we attempt at developing a method  of field
reconstruction from the integrals (\ref{RM}) and (\ref{DM}) which
allows us to avoid  these shortcomings. Our method is based on
wavelet analysis.

\section{The wavelet analysis}
\label{wl_s}

The data similar to those presented in Fig.~\ref{sampl} obviously
would not allow one to reconstruct the field in all details. What
is possible, however, is to reveal structures whose scale is
larger than the size of the gaps between the points observed. For
this purpose, one should separate large-scale structures from
noise resulting not only from statistical errors but also from the
irregular data grid. Recently wavelets have become an efficient
tool for signal analysis, especially useful in scale separation
problems. The main idea of this approach is a decomposition of the
signal over a basis of self-similar functions, known as
\emph{wavelets\/}, that are localized in both physical and Fourier
space. The wavelet transform can be considered as a generalization
of the Fourier transform, which allows the derivation of the local
spectral properties of the signal. Comprehensive reviews of the
wavelet analysis can be found in \citet{farge,gm84,holsch}.

In terms of the polar coordinates $(r,\varphi)$,
the continuous two-dimensional wavelet transform
(also known as the wavelet coefficient) of a function
$F(r,\varphi)$ is given by
\begin{equation}
w(a,r,\varphi) ={1\over\sqrt{a}}
\int\limits_{-\pi}^{\pi}\!\!\int\limits_0^\infty
\psi_{a,r,\varphi}(r',\varphi') F(r',\varphi')\, r' \,dr'\,
d\varphi', \label{wt}
\end{equation}
where $\psi_{a,r,\varphi}(r',\varphi')$ is a family of wavelets
characterized by position in the physical space ($r$,$\varphi$)
and the scale $a$ which can be identified with the radius of the
region where $\psi_{a,r,\varphi}(r',\varphi')=O(1)$. A
self-similar family of wavelets is generated from an `analyzing
wavelet' $\Psi(r,\varphi)$ by translation and dilation. Many
various wavelets have been suggested. The success of wavelet
analysis depends on appropriate choice of the wavelet
$\Psi(r,\varphi)$. Most importantly, some wavelets provide better
resolution if the wave number (Fourier) space (at the expense of
the localization information), whereas others are advantageous if
higher spatial resolution is needed (but then the scale separation
power can be only modest). The wavelet used here derives from an
isotropic wavelet known as the `Mexican hat', which has good
resolution in the physical space,
\begin{equation}
\Psi(x) = \left(2-x^2\right)\exp{\left(-\half x^2\right)}\;,
\label{mh}
\end{equation}
where $x$ is the distance from the wavelet center. An isotropic
wavelet has no preferred direction, and so rotation is not
included into the transformations generating the wavelet family,
which is defined as
\begin{equation}
\psi_{a,r,\varphi}(r',\varphi')={1\over{a}}
\Psi\left({s(r,\varphi,r',\varphi')\over{a}}\right)\;,
\label{wfam}
\end{equation}
where $s(r,\varphi,r',\varphi')$ is the distance between the
$(r,\varphi)$ (the centre of the wavelet)
and the current point $(r',\varphi')$.

Ideally, wavelet analysis consists of decomposing the signal into
contributions from a range of scales, filtering out the smaller
scales dominated by noise and the reconstruction of the filtered
signal (presumably free of noise) using the inverse wavelet
transform. However, this procedure is rarely followed completely
because this would require observational data of unrealistically
high quality. Instead, we shall consider the wavelet coefficients
themselves, but at a scale distinguished by the data.
Specifically, we focus on those scale(s) which make a dominant
contribution to the `spectral energy density' (see below). The
wavelet coefficients form a three-parametric family. For
representation and interpretation purposes, we shall usually
consider a set of two-dimensional maps of $w(a,r,\varphi)$ for a
fixed $a$.

A useful integral diagnostic of the importance of different scales
in the data is the wavelet spectrum, defined as
\begin{equation}
M(a)=\int\limits_{-\pi}^{\pi}\!\!\int\limits_0^\infty
w^2(a,r,\varphi)\,r \,dr\, d\varphi\;. \label{Ma}
\end{equation}
In the Fourier analysis, a similar quantity is known as the spectral
energy density. The normalization factor in Eq.~(\ref{wfam}) is
chosen so that structures of the same amplitude
(but different scales) contribute equally to $M(a)$.

\section{Wavelet differentiation}    \label{method_s}
In this section we suggest a general method to analyze an
observable represented by an integral, with variable limit(s), of
the physical field whose reconstruction is the goal of the
analysis. Deriving magnetic field from pulsar $\RM$ is a perfect
example of such a problem, however the method can be used in many
other problems of image analysis. The main idea of the method is
to avoid the differentiation of the discrete, noisy data. Instead,
we differentiate the wavelet itself, which is given in an
analytical form. Thus, the algorithm can be considered as a sort
of wavelet interpolation used to derive the derivative of an
observable quantity.

\subsection{The differentiating wavelet}        \label{method_s1}
Let $F(r,\varphi)$ be an unknown function whose integral
\begin{equation}
G(r,\varphi)=\int\limits_0^r F(r',\varphi) \,dr'\;, \label{fg}
\end{equation}
is known for various values of $r$. In the context of this paper,
$G(r,\varphi)$ can be $\DM$ or $\RM$ (and $F(r,\varphi)$ is
related to $\nel$ or $\nel\HH_\parallel$, respectively), and the
origin of the reference frame is at the solar position.

The exact expression for $F$ involves a derivative of $G$,
$F(r,\varphi)={\partial G(r,\varphi)}/{\partial r}$. Since $G$ is
known only at discrete data points, one has to use a smooth
interpolation of $G$ in order to derive $F(r,\varphi)$. Random and
systematic noise makes this procedure unstable. This difficulty
also arises in standard tomography problems where a
two-dimensional function is reconstructed from its integral
projections contaminated by noise \citep{pf}.

Wavelet analysis can help to alleviate this problem.
Given  Eq.~(\ref{fg}), one can rewrite Eq.~(\ref{wt}) as
\begin{equation} w(a,r,\varphi) =
{1\over{\sqrt{a}}}\!
\int\limits_{-\pi}^{\pi}\!\int\limits_0^\infty\!
\psi_{a,r,\varphi}(r',\varphi') \frac{\partial
G(r',\varphi')}{\partial r'}\; r' \,dr'\, d\varphi'. \label{2dplpr}
\end{equation}
Integrate by parts to obtain
\begin{equation}
w(a,r,\varphi) = -
{1\over{\sqrt{a}}}\int\limits_{-\pi}^{\pi}\!\!\int\limits_0^\infty
\overline{\psi}_{a,r,\varphi}(r',\varphi') G(r',\varphi')
r' \,dr'
d\varphi', \label{2dpfin}
\end{equation}
where $\overline{\psi}_{a,r,\varphi}(r',\varphi')$ is a new
wavelet family  related to the original wavelet by
\begin{equation}
\overline{\psi}_{a,r,\varphi}(r',\varphi') = \frac{\partial
\psi_{a,r,\varphi}(r',\varphi')}{\partial r'} +
\frac{\psi_{a,r,\varphi}(r',\varphi')}{r'}\;. \label{psifam}
\end{equation}
Here we have taken into account that $G(0,\varphi)=0$ and that the
wavelet is localized in the physical space,
\begin{equation}
\lim\limits_{r'\to\infty} r' \psi_{a,r,\varphi}(r',\varphi') =
0\;.
                        \label{loc}
\end{equation}

\begin{figure}
\centering
\includegraphics{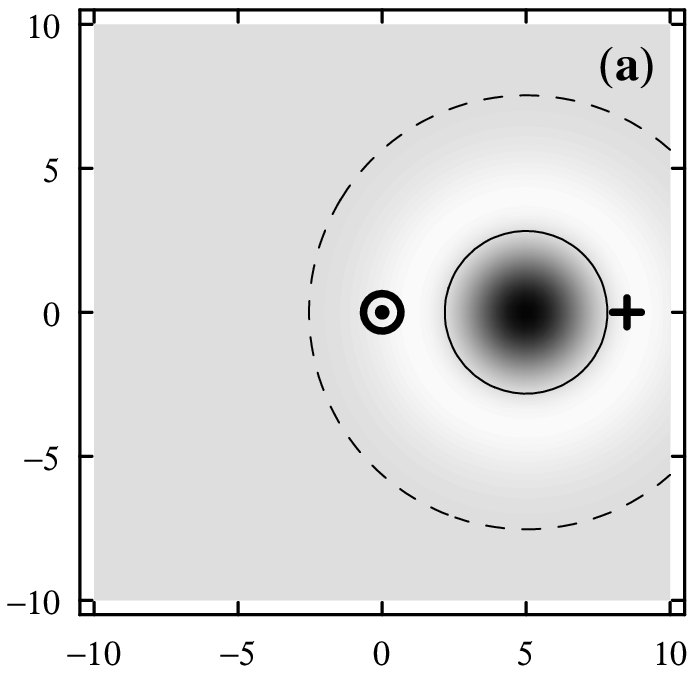}
\includegraphics{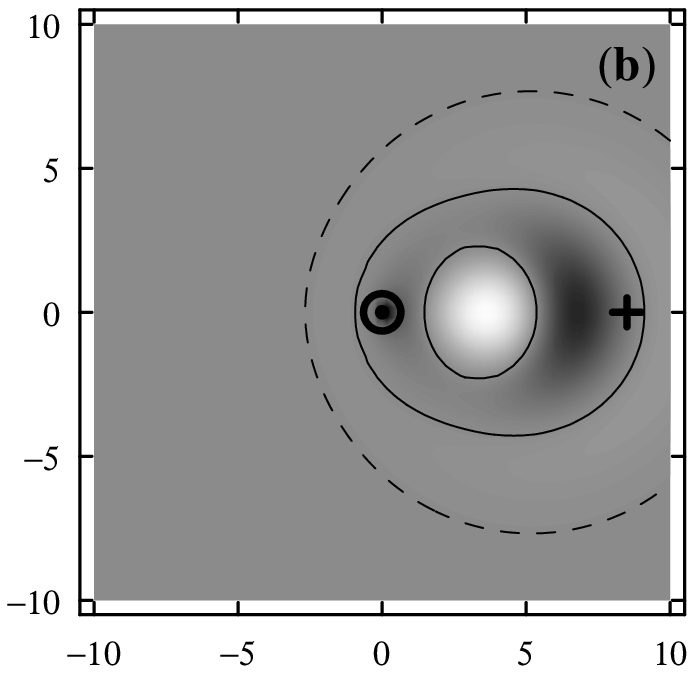}
\caption{(a) The `Mexican hat,' Eq.~(\ref{mh}), and (b) the
associated differentiating wavelet, Eq.~(\ref{psifam}). In this
illustration, the wavelets are centered at $(r,\varphi)=(5,0)$ and
have the scale $a=2$. The zero-level contours are shown solid and
the dashed contours corresponds to the magnitude of 1\% of the
maximum. Darker shades of grey correspond to larger values.}
\label{waves}
\end{figure}

Altogether, we have introduced a new family of wavelets that
allows us to reconstruct $F(r,\varphi)$ without differentiating
the data. The forms of the `Mexican hat' and the associated
differentiating wavelet are illustrated in Fig.~\ref{waves}.
Unlike the `Mexican hat', the differentiating wavelet is not
isotropic, and has two zeros.

The new wavelet $\overline\psi$ has zero mean value,
\begin{equation}
\int\limits_{-\pi}^{\pi}\!\int\limits_0^\infty
\overline{\psi}_{a,r,\varphi}(r',\varphi')\, r' \,dr' d\varphi'=0 \;,
\label{plcond1}
\end{equation}
and so satisfies what is known as the admissibility condition.
Another admissibility condition is that
\begin{equation}
\int\limits_{-\pi}^{\pi}\!\int\limits_0^\infty
\overline{\psi}_{a,r,\varphi}(r',\varphi')\, r'^2 \,dr' d\varphi'=0 \;.
\label{plcond2}
\end{equation}
Equations (\ref{plcond1}) and (\ref{plcond2}) imply that the
wavelet coefficients are independent of a constant component of
$G(r,\varphi)$ and its linear trend, respectively, -- a behavior
required for any wavelet in order to have an inverse
transformation. Thus, $F(r,\varphi)$ can be determined up to a
constant background value proportional to the value of
$\HH_\parallel$ at $r=r_{\rm max}$, where $r_{\rm max}$ is the
distance to the most remote pulsar. If $r_{\rm max}$ is large
enough, this background value  is close to zero.

\subsection{Implementation for irregularly spaced data}
\label{method_s2}

The calculation of the integral (\ref{2dpfin}) on a discrete,
irregular data grid is not straightforward. The need for a robust
numerical algorithm is especially demanding since the separation
of pulsars in the available catalogues is often comparable to
or exceeds the physically interesting scales of the large-scale
Galactic magnetic field.  The difficulties becomes crucial if the
separation of the data points can be comparable to the scale of the
wavelet, $a$.

In our case, a convenient method to calculate the integrals
is the Monte-Carlo technique, leading to
\begin{equation}
w(a,r,\varphi) = - {\Delta
S(a,r,\varphi)\over{\sqrt{a}}}\sum_{i=1}^N
\overline{\psi}_{a,r,\varphi}(r'_i,\varphi'_i)\,
G(r'_i,\varphi'_i)\, r'_i \,, \label{2dpfind}
\end{equation}
where $N$ is the number of data points for $G(r,\varphi)$ and
$\Delta S$ is the average area element of an individual data
point. For a uniform distribution of the data points, $\Delta S$
is a constant, but this is not true for a non-homogeneous
distribution. We chose $\Delta S$ (a function of position and the
scale of the wavelet, identified with the effective weight of data
points) as to minimize the error arising from the non-uniform
distribution of the data points. The contribution of a data point
to the integral sum (\ref{2dpfind}) must be small if the point is
far away from the centre of the wavelet, $\vec{r}'$, as compared
with the scale of the wavelet, $a$, so that we take
\begin{equation}
\Delta S(a,r,\varphi)=
a^2\left[{\sum_{i=1}^N
\Phi\left({s(r,\varphi,r_i,\varphi_i)\over{a}}\right)}\right]^{-1}\,.
\label{ds}
\end{equation}

The value of $a^2/\Delta S$ gives the approximate number of
sample points within  the wavelet localization area. So,
$a^2/\Delta S$ should be large enough in order to achieve
sufficient accuracy.

Another problem arising in the numerical realization of the
wavelet transformation is the necessity to satisfy the
admissibility conditions (\ref{plcond1}) and (\ref{plcond2}) with
sufficient accuracy. We use for this purpose the gapped (or
adaptive) wavelet technique \citep[see][]{frgr} wherein the
`analyzing wavelet' is represented in the form
\begin{equation}
\Psi(x) = h(x)  \Phi(x)\;,      \label{w_a}
\end{equation}
where $\Phi(x)$  is a positive definite function that determines
the position of the wavelet (the envelope). For the `Mexican hat',
Eq.~(\ref{mh}), $\Phi(x)=\exp(-\half x^2)$ and $h(x)=2-x^2$. The
main idea of the  technique is to modify $h(x)$ in order to
satisfy the two conditions (\ref{plcond1}) and (\ref{plcond2}).
This can be achieved by introducing two quantities
 $C_1$ and $C_2$, functions of
scale $a$ and  position $(r,\varphi)$, such that
\begin{equation}
\Psi(x) = [h(x)+C_0+C_1 x]  \Phi(x)\;. \label{w_ap}
\end{equation}
These functions  are  determined from the discrete versions of
Eqs.~(\ref{plcond1}) and (\ref{plcond2})  for each scale $a$ and
wavelet position $(r,\varphi)$.

\section{Testing the method}
\label{demo_s}

In order to assess the possibilities of the method, we consider a
test function $F(r,\varphi)=\nel$ similar to that of  the Milky
Way \citet{1993ApJ...411..674T}. We first calculate the
corresponding values of $G(r,\varphi)\propto\DM(r,\varphi)$ using
Eq.~(\ref{RM}) and then apply our method to obtain $F(r,\varphi)$
from $G(r,\varphi)$ on a selection of data grids.

\begin{figure} \centering \includegraphics{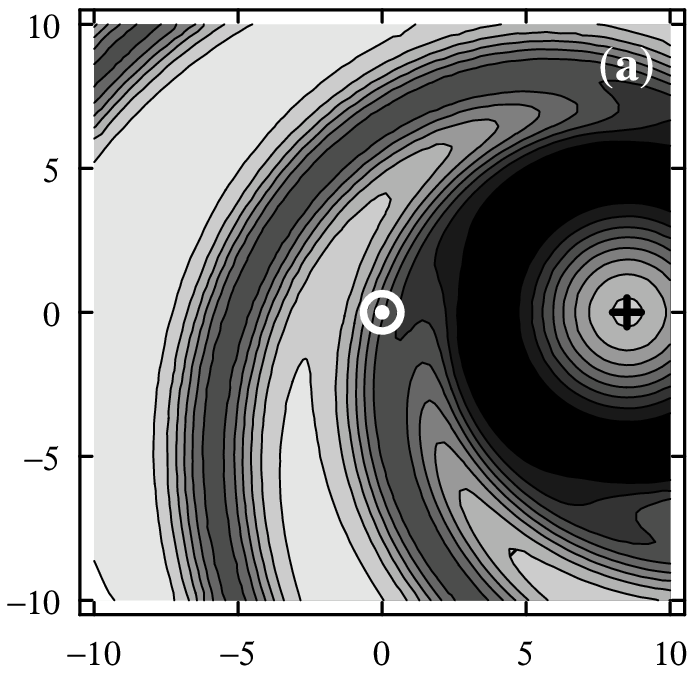}
\includegraphics{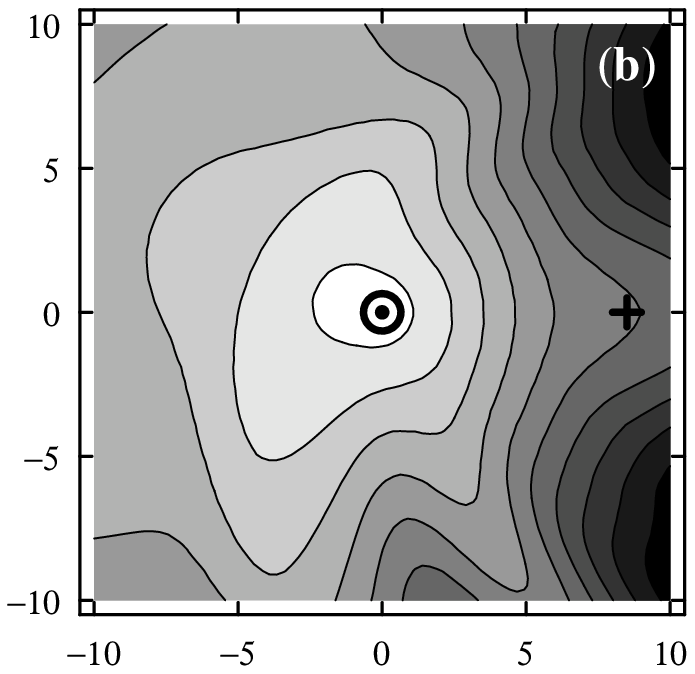}
\caption{(a) The test function $F$ and (b) the corresponding
distribution of $G$ calculated from $F$ using Eq.~(\ref{fg}). With
$F=\nel$, $G$ is proportional to $\DM$. Lighter shades of grey
correspond to smaller values.} \label{plsig}
\end{figure}

We show in the upper panel of Fig.~\ref{plsig} a test function
$F=\nel$ that includes an annular and spiral structures with
characteristic scales (transverse half-widths)
$a=3\,$kpc and $a=1\,$kpc, respectively. This distribution in
similar to (but not identical with) that of Fig.~\ref{ne}. The
lower panel of Fig.~\ref{plsig} shows the corresponding
$G=\DM(r,\varphi)$ calculated in the whole plane.

The quality of the reconstruction of $F$ from $G$ obviously
depends on how well $G$ is sampled. For a reliable reconstruction
of a simple, isotropic isolated object whose structure is similar
to that of the wavelet itself, one needs at least 10 data points
within it. Then one would need about 2000 data points distributed
uniformly in $(r,\varphi)$ in order to detect structures of the
smallest scale of 1\,kpc in the
$20\,\mbox{kpc}\times20\,\mbox{kpc}$ Solar vicinity.  However,
Faraday rotation measures are only known for about 300 pulsars;
furthermore, their distribution is very non-uniform being
 very sparse beyond
3--5\,kpc from the Sun, especially in the fourth Galactic
quadrant. Thus, the available pulsar data can only yield reliable
information about structures in a narrower vicinity of the Sun
and/or at a scale of a few kiloparsecs.

Now we consider a series of wavelet transforms on different data
grids, from an `ideal' to a realistic case through an `acceptable'
one.

\begin{figure}
\centering
\includegraphics{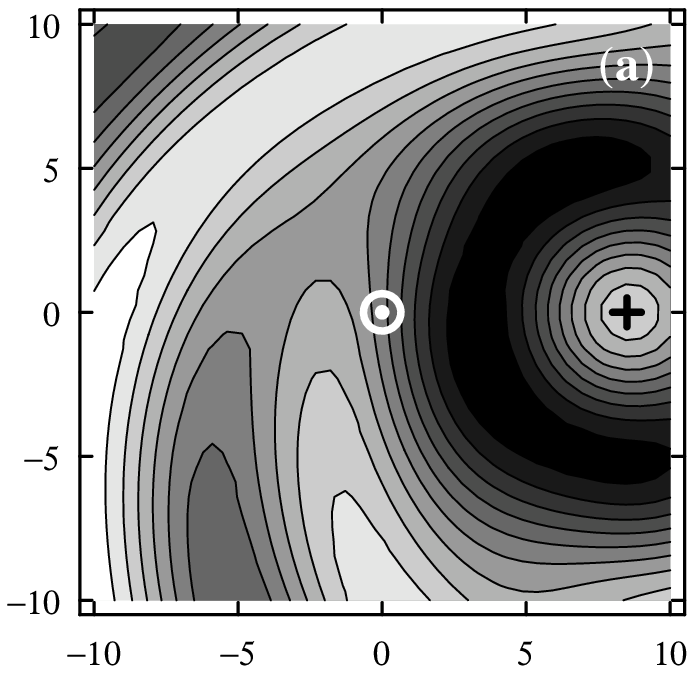}
\includegraphics{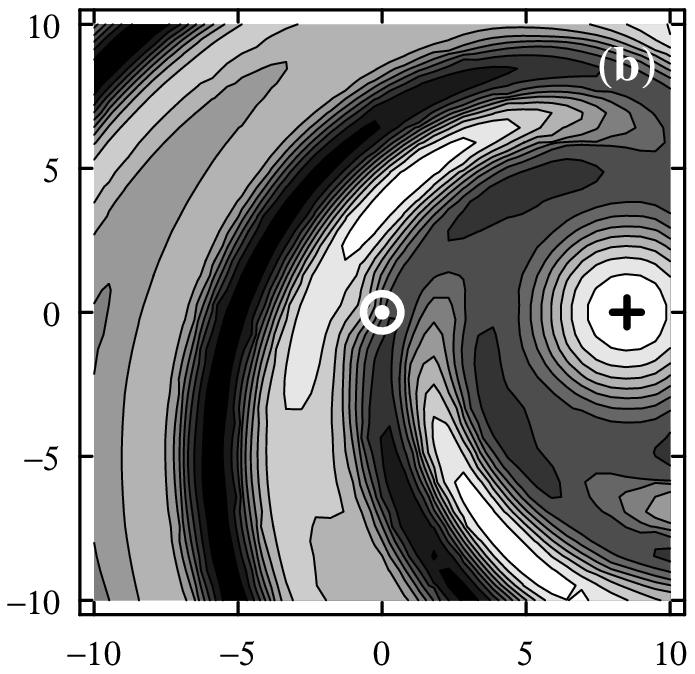}
\caption{The wavelet transform of the distribution of
Fig.~\ref{plsig}b on a regular $40\times 40$ grid at a scale (a)
$a=3\,$kpc (the scale of the annulus) and (b) $a=1\,$kpc (the
scale of the spirals). } \label{pltest1}
\end{figure}

\begin{figure}
\centering
\includegraphics[width=0.85\columnwidth]{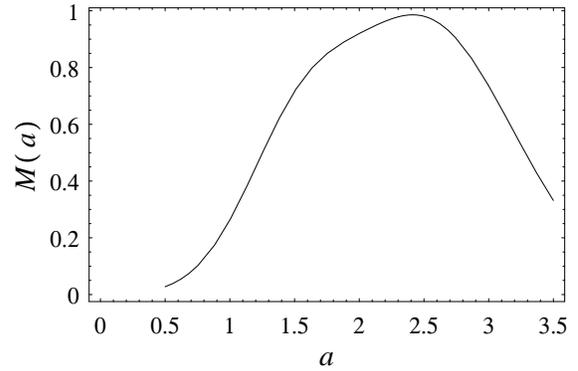}
\caption{The integral wavelet spectrum $M(a)$ for the distribution
of Fig.~\ref{plsig}b, sampled on a regular $40\times40$ grid, with
$a$ in kpc.} \label{spectrum}
\end{figure}

A regular grid is the best for the calculation of the integral
(\ref{2dpfin}). The data point separation of $0.5\,$kpc (needed to
recover structures at a scale of 1\,kpc) requires a sample of 1600
points. The wavelet transforms of the distribution of
Fig.~\ref{plsig} sampled on a regular grid of 1600 points are
shown in Fig.~\ref{pltest1} for the scales $a=3\,$kpc and
$1\,$kpc. As expected, the annular structure is better pronounced
at $a=3\,$kpc (Fig.~\ref{pltest1}a) and the spirals are better
visible at $a=1\,$kpc (Fig.~\ref{pltest1}b). Figures
\ref{pltest1}a and b together reproduce the distribution of
Fig.~\ref{plsig} reasonably well, demonstrating the possibilities
of the wavelet transform in scale separation. However, the scale
resolution of the wavelet, derived from the `Mexican hat', is only
modest, and this contaminates the wavelet transform at $a=3\,$kpc
with a trace of the spiral arms, and that at $a=1\,$kpc with a
weak signature of the annulus.  The magnitude of
the wavelet coefficient is maximum when the scale of the wavelet
is equal to the scale of the structure. The integral wavelet
spectrum $M(a)$, defined in Eq.~(\ref{Ma}) and shown in
Fig.~\ref{spectrum} confirms that the scales $1\le a\le3$ are
dominant in $G(r,\varphi)$, but they are not separated from each
other.

\begin{figure}
\centering
\includegraphics{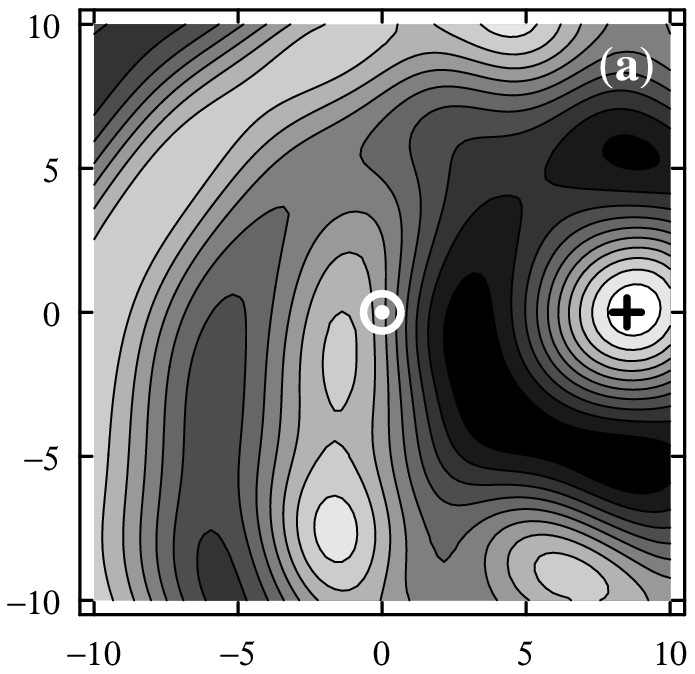}~(a)
\includegraphics{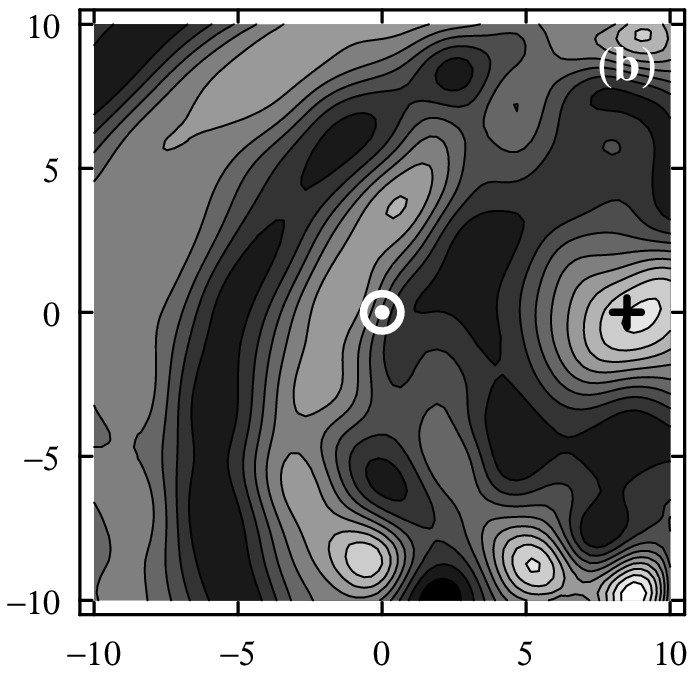}~(b)
\caption{The wavelet transforms of the distribution of
Fig.~\ref{plsig}b sampled at 1600 randomly distributed points at
the scales (a) $a=3\,$kpc and (b) $a=1\,$kpc.} \label{pltest2}
\end{figure}

\begin{figure}
\centering
\includegraphics{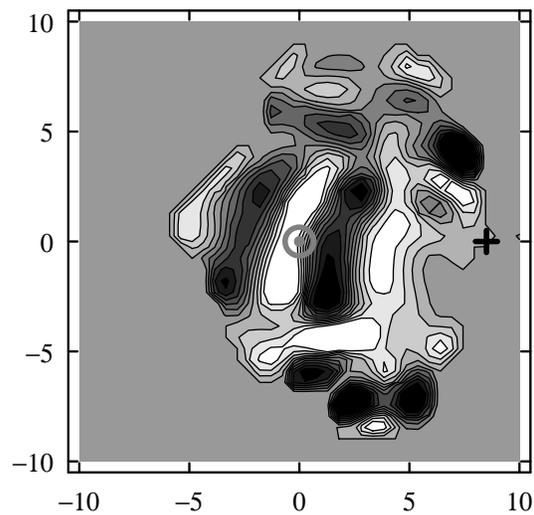}
\caption{The wavelet transform of the distribution
of Fig.~\ref{plsig}b on the real, irregular data grid
of the pulsar catalogue of \citet{tay95} (323 points)
at a scale $a=1.5\,$kpc. The region where $\Delta S>a^2/5$
and the results are inaccurate is filled with uniform grey.}
\label{pltest3}
\end{figure}

Further, we consider the data of Fig.~\ref{plsig}b sampled at the
same number of points (1600), but now the data points are
scattered randomly and statistically uniformly in the same region
$20\,\mbox{kpc}\times20\,\mbox{kpc}$ around the Sun. The
distortions in the resulting wavelet transform maps of
Fig.~\ref{pltest2} are significant even though the two basic
components of the distribution are clearly recognizable. The
structures in the wavelet transform become patchy, especially at
small scales, because of the large gaps in the data grid.

Finally, the wavelet transform on the data grid of the pulsar
catalogue of \citet{tay95} (using only the positions of 323
pulsars with known $\RM$) is shown in Fig.~\ref{pltest3}. The data
points are crowded within 3\,kpc from the Sun. We consider the
wavelet transform to be unreliable where $\Delta S>a^2/5$; these
regions are marked with uniform grey shade in Fig.~\ref{pltest3}.
The spiral arms segments of the original distribution of
Fig.~\ref{plsig}a is recognizable in the 3\,kpc vicinity of the
Sun, but they appear patchy and discontinuous. The annulus at the
galactocentric distance of 4\,kpc is hardly visible at all, and
the arm-interarm contrast is overestimated.

\section{Discussion}
\label{diss_s} We have introduced a new wavelet devised for the
analysis of the pulsar Faraday rotation measures in terms of the
large-scale magnetic field (or of any other observable that is an
integral of the quantity studied, e.g., the pulsar dispersion
measures). This is a tomography approach because the field is
directly reconstructed from its integral estimator. The method
works well with data given on a regular mesh or scattered randomly
but with gaps between the data points not exceeding $a/2$, with
$a$ the  scale of the wavelet. The separation of pulsars with
known $\RM$ exceeds this limit beyond about 3\,kpc from the Sun.

An advantage of the method is that it  involves neither {\it ad
hoc\/} assumptions about magnetic field structure nor model
fitting. However, the wavelet transforms on the data grid of the
pulsar catalogues available appear rather confusing and difficult
to interpret.  The advantages of the wavelet analysis and model
fitting can be combined in a single approach applied by \cite{mw}
to the Faraday rotation measures of extragalactic sources. Instead
of fitting a magnetic field model to noisy $\RM$ data, these
authors fitted the wavelet transform of the model $\RM$ to the
wavelet transform of the observed $\RM$ with smaller scales
(mainly responsible for the nose) filtered out. This has resulted
in a significant improvement in the quality of the analysis. The
application of the wavelet introduced here will allow us to fit
the wavelet transforms of the magnetic field derived from $\RM$ to
the model. We expect that will result in a significant improvement
of the results against the usual procedure of fitting model $\RM$
to the observed noisy data.

Another way to improve the results would be to analyze
simultaneously the Faraday rotation measures of pulsars and
extragalactic radio sources. Since many extragalactic sources
occur at high Galactic latitudes, a plausible assumption about the
vertical distribution of the magneto-ionic medium will have to be
adopted.

\begin{acknowledgements}
We acknowledge financial support from the Russian Foundation for
Basic Research (Grant 01-02-16158), NATO Collaborative Linkage
Grant PST.CLG~974737 and the University of Newcastle (Small Grants
Panel). RS thanks DAAD for support and Astrophysikalisches
Institut Potsdam for hospitality.
\end{acknowledgements}

\bibliographystyle{aa}
\bibliography{wtpaper}

\end{document}